\documentclass[prl,twocolumn,showpacs,preprintnumbers,amsmath,amssymb]{revtex4}

\usepackage{graphicx}
\usepackage{dcolumn}
\usepackage{bm}

\begin{document}


\title{Adiabatic spin pumping through a  quantum dot with a single orbital level}

\author{Tomosuke Aono}
\email{aono@ph.sci.toho-u.ac.jp}
\affiliation{%
Department of Physics, Toho University\\
2-2-1 Miyama, Funabashi, Chiba 274-8510 JAPAN, and
Department of Physics, Ben-Guiron University\\
Beer-Sheva 84105, ISRAEL
}%

\date{\today}

\begin{abstract}
We investigate an adiabatic spin pumping
through a quantum dot with
a single orbital energy level under
the Zeeman effect.
Electron pumping is produced by
two periodic time dependent parameters,
a magnetic field and
a difference of the dot-lead coupling between
the left and right barriers of the dot.
The maximum charge transfer per cycle is 
found to be $e$, the unit charge
in the absence of a localized moment in the dot.
Pumped charge and spin are different,
and spin pumping is possible without charge pumping in
a certain situation.
They are
tunable by changing 
the minimum and maximum value of the magnetic field.
\end{abstract}

\maketitle

Adiabatic electron pumping is the mechanism
which produces
a finite charge transfer through a system
when the system is altered slowly
by external parameters and
it is returned to its initial state after a certain period~\cite{Thouless83}.
In quantum dot systems,
electron pumping has been realized
in electron turnstile~\cite{Kouwenhoven91,Pothier92,Keller98},
through which
a quantized charge is transferred per cycle
under controlled gate voltages
by the Coulomb blockade effect.
In a recent experiment~\cite{Switkes99},
an adiabatic quantum electron pumping is realized
in an open quantum dot 
under two oscillating gate voltages and zero-source drain voltage,
where
electrons are transferred by
electron interference effect through the system.
Many theoretical works have been published in relation to this pumping
\cite{Spivak95,Aleiner98,Brouwer98,Brouwer98_2,Shutenko00,Zhou99,Avron00,
Polianski01,Aleiner00,Levinson00,Moskalets01,Wei00,Cremers01,Simon00,Sharma01,Mucciolo01,Entin02}.

Under magnetic fields,
adiabatic spin pumping will occur,
which is proposed in
the Tomonaga-Luttinger liquid~\cite{Sharma01} and
an open quantum dot~\cite{Mucciolo01}.
Spin pumping will be very useful in
developing of spin dependent transport,
especially spin injection methods.
Spin injection into semiconductor materials
is realized
using ferromagnetic metals~\cite{Lee99,Hammar99}
or magnetic semiconductor~\cite{Fiederling99,Ohno99}
contacts.
These injections are driven by
chemical potential differences
across the samples.
In contrast,
spin pumping works under zero source-drain voltage
without magnetic materials.
We propose an adiabatic spin pumping
through a quantum dot with a single orbital energy level
by using an oscillating magnetic field.
First we investigate the magnitude of electron pumping and
then the separation of pumped charge and spin,
including 
spin pumping without charge pumping
as an extreme case.

\begin{figure}
\includegraphics[width=0.25\textwidth]{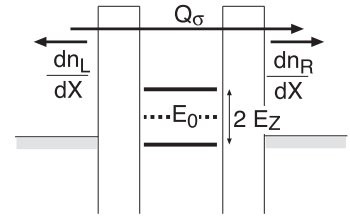}
\caption{\label{fig:dot} 
Schematic view of a  quantum dot with a single orbital energy level
under a magnetic field.
}
\end{figure}

We consider a quantum dot system as shown in Fig.~\ref{fig:dot}.
The dot has
a single dot level $E_0$  
and couples to the two leads $\alpha = L,R$ with
a tunneling matrix element $T_{\alpha}$.
The tunneling coupling results in a level broadening
of the dot level $\Gamma = \Gamma_L + \Gamma_R$, where
$\Gamma_{\alpha} = \pi \rho | T_{\alpha} |^2$
with the density of states $\rho$ at the Fermi level in
the leads.
When a magnetic field $B$ is applied,
the dot level splits into
$E_{\sigma} = E_0 - \sigma E_Z $
with the Zeeman energy, $E_Z = g/2 \;\mu_B B$
and the spin index $\sigma$ ($\sigma=\pm$).
We assume the Zeeman effect is negligible in the leads~\cite{Zeemaninlead}.
The Coulomb interaction in the dot is taken into account.
To describe this system,
we adopt the Anderson model 
\cite{Glazman88,Ng88}:
\begin{eqnarray}\label{eq:Hamiltonian}
H &=&
\sum_{k,\sigma,\alpha=L,R} 
 \epsilon_{k} c^{\dagger}_{k\sigma\alpha} c_{k\sigma\alpha}+
\sum_{\sigma=\pm}
  E_{\sigma} d^{\dagger}_{\sigma} d_{\sigma}
+U n_{+} n_{-} \nonumber\\
&& + \sum_{k,\sigma,\alpha=L,R}
 \left( T_{\alpha} c^{\dagger}_{k\sigma\alpha}
d_{\sigma} +
\textrm{h.c.}
\right).
\end{eqnarray}
Here
$c^{\dag}_{k \sigma \alpha}$ creates an electron
with
energy $\epsilon_{k}$ and spin $\sigma$
in lead  $\alpha = L, R$,
$d^{\dag}_{\sigma}$ creates an electron in
the dot with spin $\sigma$,
$n_{\sigma} = d^{\dagger}_{\sigma} d_{\sigma}$, and
$U$ is the strength of the Coulomb interaction.
The Hamiltonian (\ref{eq:Hamiltonian}) is reduced 
to the single impurity Anderson model~\cite{Anderson61}
by a unitary transformation for
electrons in the leads:
$c_{k \sigma} = u^* c_{k \sigma L} + v^* c_{k \sigma R}$,
and $\bar{c}_{k\sigma} = -v c_{k \sigma L} + u c_{k \sigma R}$
with  $u=T_{L}/T$, $v=T_{R}/T$ and $T = \sqrt{|T_{L}|^2 + |T_{R}|^2}$
~\cite{Glazman88}.
After the transformation,
the modes $\bar{c}_{k\sigma}$ do not couple with electrons in the dot while
the modes $c_{k\sigma}$ couple with $d^{\dagger}_{\sigma}$
through the tunneling matrix element $T$.

To this system,
we apply adiabatic external sources
with a frequency $\omega$.
We assume
$\omega \ll \Gamma$ 
while $\omega$ is much larger than
the Kondo temperature.
The former guarantees
the adiabatic condition, and
the latter means
the Kondo effect is suppressed~\cite{Kaminski99}.
Accordingly we disregard 
spin exchange mechanisms and
regard spin as a good quantum number.
For simplicity,
we discuss the zero temperature limit.

In general,
an adiabatic pumping requires two periodic external parameters,
$X_1$ and $X_2$,
with the common frequency $\omega$.
Pumped electron charge $Q_{\sigma}$ with spin $\sigma$
from the left to right lead
after a period $\tau=2\pi/\omega$ is given by 
\begin{equation}
   Q_{\sigma} = -e\int_0^\tau dt \left( \frac{dn_{L,\sigma}}{dX_1} \frac{dX_1}{dt}
     + \frac{dn_{L,\sigma}}{dX_2} \frac{dX_2}{dt} \right).
\label{eq:adiabatic_charge}
\end{equation}
Here $dn_{\alpha,\sigma}/dX $ is the emissivity
into the lead $\alpha\;(\alpha=L,R)$,
which is
the number of electrons with spin $\sigma$ entering into the lead $\alpha$
as a result of
the charge redistribution caused by $X$.
It is expressed in terms of
the matrix elements of
the scattering matrix of the dot, $S_{\sigma;\alpha \beta}$ ($\alpha,\beta=L,R$)~\cite{Buttiker94}:
\begin{equation}
   \frac{dn_{\alpha,\sigma}}{dX} = -\frac{1}{2\pi} \sum_{\beta=L,R}
\textrm{Im} 
\left[
S^*_{\sigma;\alpha \beta}
 \frac{\partial S_{\sigma;\alpha \beta}}{\partial X}
 \right].
\label{eq:adiabatic_number}
\end{equation}
Then Eq.~(\ref{eq:adiabatic_charge})
is expressed by a two dimensional integral~\cite{Brouwer98}:
\begin{equation}
Q_{\sigma} = e \int \int dX_1 dX_2 \Pi_{\sigma}(X_1,X_2)
\end{equation}
 with 
\begin{equation}\label{eq:pump_charge_Green}
\Pi_{\sigma}(X_1,X_2) =\frac{1}{\pi}
{\rm Im} \left[ 
\frac{\partial S^*_{\sigma; LL} }{\partial X_1} 
\frac{\partial  S_{\sigma; LL}}{\partial X_2} +
\frac{\partial S^*_{\sigma; LR} }{\partial X_1} 
\frac{\partial  S_{\sigma; LR}}{\partial X_2} 
 \right].
\end{equation}
A current $I_{\sigma}$ with spin $\sigma$ is given by
$I_{\sigma} = \omega Q_{\sigma}/2 \pi$.

To investigate an adiabatic spin pumping through the dot system,
we choose a set of pumping parameters;
one is the Zeeman energy, $E_Z(t)$, and
the other is the asymmetry factor $p(t)$
defined by
$p(t) = (\Gamma_L - \Gamma_R)/\Gamma$,
while $\Gamma_L + \Gamma_R$ is kept at a constant value $\Gamma$.
From its definition, $ -1 \le p(t) \le 1$.
In general,
the matrix elements of the scattering matrix $S_{\sigma} (X_1,X_2)$ through the dot are given
by \cite{Ng88}:
\begin{eqnarray}\label{eq:S_matrix_dot}
S_{\sigma; LL/RR}(X_1,X_2) &=&  1- 2 i \Gamma_{L/R} G_{\sigma},\nonumber\\
S_{\sigma; LR/RL}(X_1,X_2) &=&  -2i \sqrt{\Gamma_L \Gamma_R} e^{\pm i \gamma}
G_{\sigma},
\end{eqnarray}
where $G_{\sigma}$ is the single-particle Green function for an electron in the dot with spin
$\sigma$ at the Fermi level of the leads, and
$\gamma = {\rm Arg} (T_L T^*_R)$,
which does not appear explicitly in
the following discussion.
Substituting Eq.~(\ref{eq:S_matrix_dot}) into Eq.~(\ref{eq:pump_charge_Green}) yields
$Q_{\sigma} = e \int \int  dp \;dE_Z \;\Pi_{\sigma}(E_Z,p)$ with 
\begin{equation}\label{eq:charge_Green}
\Pi_{\sigma}(E_Z,p)   = \frac{1}{\pi} {\rm Im} \left[ \frac{\partial G_{\sigma}^{*}}{\partial E_Z}
G_{\sigma} \right].
\end{equation}
Since $G_{\sigma}$ is the Green function of
the single impurity Anderson model~\cite{Anderson61},
it is independent of $p$.
This results in
$\Pi_{\sigma} (E_Z,p) = \Pi_{\sigma}(E_Z)$.
Then the integration over $p$ in the expression of $Q_{\sigma}$ is replaced by
a constant value; 
in the following, it is equals to $-2$.

First we investigate the magnitude of electron pumping.
For this purpose,
$Q_{\sigma}$ is re-expressed by the integration of $p$ and
the Friedel phase $\delta_{\sigma}$, which is the phase of
the transmission coefficient through the dot.
It determines the transmission probability $T_{\sigma}$  through the dot for electrons with spin
$\sigma$:
\begin{equation}\label{eq:transmission}
T_{\sigma} = ( 1- p^2) \sin^2 \delta_{\sigma}.
\end{equation}
Furthermore, it satisfies the Friedel sum rule\cite{Langer61,Langreth66}:
\begin{equation}\label{eq:Friedel_Sum_Rule}
\langle n_{\sigma} \rangle = \delta_{\sigma}/\pi
\end{equation}
with the occupation number $\langle n_{\sigma} \rangle = \langle  d^{\dagger}_{\sigma} d_{\sigma}
\rangle$.
Since
$\langle n_{\sigma} \rangle$ is a function of $E_Z$,
$\delta_{\sigma} = \delta_{\sigma}(E_Z)$.
Using
$\exp (2 i \delta_{\sigma}) = 1 - 2 i \Gamma G_{\sigma}$~\cite{Langer61,Langreth66}
and substituting it into Eq.~(\ref{eq:charge_Green}),
we finally obtain
\begin{equation}\label{eq:charge_Friedel}
Q_{\sigma} =\frac{2e}{\pi} \int_{\delta_{\sigma,1}}^{\delta_{\sigma, 2}} 
d\delta_{\sigma}\;\sin^2 \delta_{\sigma},
\end{equation}
where we have integrated over $p$ and introduced the lower (upper) limit of the integration,
$\delta_{\sigma,1 (2)}$.
Equation (\ref{eq:charge_Friedel}) shows that
$Q_{\sigma}$ has a maximum value $e$ 
when $\delta_{2,\sigma} = \pi$
and
$\delta_{1,\sigma}=0$,
where $\delta_{\sigma}$
changes so as to run through the resonance of $T_{\sigma}$.
[See  Eq.~(\ref{eq:transmission}).]
This result coincides with
the condition of the maximum pumping
~\cite{Levinson00,Entin02},
which states that it occurs when
the trajectory of the pumping parameters encircle the peak of the transmission probability.
Note that we have shown this condition in the presence of the Coulomb interactions.

Until now we have assumed
the adiabatic condition is satisfied during the pumping cycle.
This is, however, not obvious 
if a localized moment appears in the dot.
This is because a magnetization of the localized moment responds sensitively to
the change of sign of an applied magnetic field.
Thus the time dependence of the magnetization may
breaks the adiabatic condition.
Accordingly, 
it will make a difference in electron pumping
whether the local moment in the dot appears or not.
We  discuss this point in the following.

\begin{figure}
\includegraphics[width=0.5\textwidth]{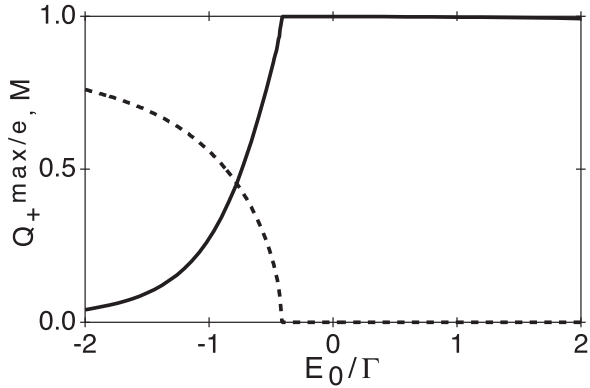}
\caption{\label{fig:decoherence} 
The maximum value of up-spin pumped charge $Q_{+}^{\rm max}$ (solid line) and
$M = \langle n_{+}(0^{+}) \rangle -\langle n_{-}(0^{+}) \rangle$
(broken line) as a function of $E_0/\Gamma$ when $U/\Gamma=10$.}
\end{figure}

For this purpose,
we calculate $Q_{\sigma}$ by performing the mean field approximation
for the Coulomb interaction term:
\begin{equation}\label{eq:MFA}
U n_{+} n_{-} = 
\sum_{\sigma}  U \langle n_{-\sigma} \rangle  d^{\dagger}_{\sigma} d_{\sigma}
-U \langle n_{+} \rangle  \langle n_{-} \rangle,
\end{equation}
where $\langle n_{\sigma} \rangle=\langle n_{\sigma}(E_Z) \rangle$ is determined by
the self-consistent equations,
$
\cot(\pi \langle n_{\sigma} \rangle) = \left( E_{\sigma} + U \langle
n_{-\sigma} \rangle \right)/\Gamma.
$
Using this approximation,
we can qualitatively understand
how the localized moment appears in the dot,
characterized by
$M \equiv \langle n_{+}(0^{+}) \rangle- \langle n_{-}(0^{+}) \rangle$,
determined by
the value of $E_0/\Gamma$ for a fixed value of $U/\Gamma$~\cite{Anderson61}.
In Fig.~\ref{fig:decoherence}, $M$ is plotted as a function of $E_0/\Gamma$ when 
$U/\Gamma=10$
with a broken line.
When $E_0$ is large, $M=0$.
At a certain negative value of $E_0$,
$M$ starts to increase.

After this approximation,
$Q_{\sigma}$ is given by
$Q_{\sigma} = e \int \int  dp \;dE_Z \;\Pi_{\sigma}(E_Z)$
with
\begin{equation}\label{eq:pump_spin}
\Pi_{\sigma}(E_Z) = \frac{-1}{\pi} 
\frac{\Gamma^3 \left(  \sigma - U\frac{\partial \langle n_{-\sigma}
\rangle}{\partial E_Z} 
\right)} {\left[ \left( E_{\sigma}+U \langle n_{-\sigma} \rangle \right)^2 +\Gamma^2
\right]^2},
\end{equation}
where we have used that
$
G_{\sigma} = -1/(E_{\sigma}+ U \langle n_{-\sigma} \rangle - i \Gamma).
$
In Fig.~\ref{fig:decoherence},
the maximum of $Q_{+}$, $Q_{+}^{\rm max} \equiv e \int_{-\infty}^{\infty} 
d E_Z \Pi_{+}(E_Z)$ is plotted as a function of $E_0/\Gamma$
with a solid line.
When $M=0$,
$Q_{+}^{\rm max}=e$.
As $M$ increases from zero,
$Q_{+}^{\rm max}$
decreases monotonically.

Now we interpret this result
according to the Friedel phase shift argument.
If we apply $-E_{Z} \le E_Z(t) \le E_{Z}$,
the phase shift, $\Delta \delta =\delta_{+,2}-\delta_{+,1}$ is given by
\begin{equation}\label{eq:phase_shift_M}
\Delta \delta = \int_{\delta_{+}(-E_{Z})}^{\delta_{+}(0^-)} d \delta_{+} +
\int_{\delta_{+}(0^+)}^{\delta_{+}(E_Z)} d
\delta_{+} = \pi ( M(E_Z)- M )
\end{equation}
with the magnetization $M(E_Z) = \langle n_{+}(E_Z) \rangle - \langle n_{-}(E_Z) \rangle$.
The rightmost relation follows from
the Friedel sum rule (\ref{eq:Friedel_Sum_Rule}) and 
$\langle n_{+}(-E_Z) \rangle  = \langle n_{-}(E_Z)\rangle$.
Equation~(\ref{eq:phase_shift_M}) means
the maximum value of the phase shift is less than $\pi$ when $M \neq 0$.
On the other hand, 
the maximum value of $Q_{+}$ occurs only when $\Delta \delta=\pi$
since
the integrand of Eq.~(\ref{eq:charge_Friedel}) is
a positive definite function.
Hence
$Q_{+}^{\rm max}$ is suppressed when $M\neq0$.

Appearance of $M$,
for the dot system under time dependent magnetic fields,
is the reflection of the rapid change of $M(E_Z)$ around $E_Z=0$. 
In this region,
we cannot treat it as an adiabatic variable but rather
treat it only as the variable averaged over a certain interval of time
because of its rapid change,
the value of which is here represented by $M$.
Consequently, the Friedel phase shift in the rapid change region is
also replaced by
a certain averaged value.
In recent theories~\cite{Moskalets01,Cremers01},
it has been shown that
to see adiabatic pumping clearly,
electron coherence throughout the dot is essential and
electron pumping is suppressed when the coherence is broken.
In the present model,
the source of decoherence is
the rapid change of the localized moment.

Next, we consider the contributions from
both of the up and spins to
discuss the separation of pumped charge and spin.
First we note 
an antisymmetric relation between
$\Pi_{+}(E_Z)$ and $\Pi_{-}(E_Z)$:
\begin{equation}\label{eq:antisym}
\Pi_{+}(E_Z) = - \Pi_{-}(-E_Z),
\end{equation}
which is shown as follows:
Since
$\langle n_{+} (E_Z) \rangle = \langle n_{-}(-E_Z) \rangle$,
\begin{equation}
S_{+} (E_Z,p) = S_{-} (-E_Z,p),
\end{equation}
yielding
\begin{equation}
\frac{\partial S_{+} (X_1,p)}{\partial X_1} \Biggl|_{X_1=E_{Z}}=
- \frac{\partial S_{-} (X_1,p)}{\partial X_1} \Biggl|_{X_1=-E_{Z}}.
\end{equation}
Substituting this relation into Eq.~(\ref{eq:pump_charge_Green}) gives
Eq.~(\ref{eq:antisym}).
Equation (\ref{eq:antisym}) indicates
the pumped charge, $Q_{\rm charge}=Q_{+} + Q_{-}$, and
the pumped spin, $Q_{\rm spin}=Q_{+} - Q_{-}$ are different in general,
determined by the minimum and maximum values of $E_Z$, $E_{Z}^{\rm min}$ and $E_{Z}^{\rm max}$.

More importantly, Eq. (\ref{eq:antisym}) means that both of up-spin and down-spin pumped
charge
tend to flow in the opposite directions.
This leads the spin pumping without the charge pumping,
$Q_{\rm spin} \neq 0$ and $Q_{\rm charge}=0$,
when $E_{Z}^{\rm min}= -E_{Z}^{\rm max}$.
There are a few proposals of spin pumping\cite{Sharma01,Mucciolo01}.
In the present system,
the spin pumping without charge pumping
has the following properties.
First, it is always achieved
independent of the magnitude of $E_Z$ if
$E_{Z}^{\rm max}=-E_{Z}^{\rm min}\neq 0$.
Second, it simply comes from the fact that 
up and down spins have the opposite signs of the Zeeman energy,
not from the presence of electron-electron interactions.
These properties will be maintained even for
other choices of $X_2$ instead of $p$ because
the asymmetric property similar to
Eq.(\ref{eq:antisym}) is always valid:
$\Pi_{+}(E_Z,X_2) = - \Pi_{-}(-E_Z,X_2)$.

\begin{figure}
\includegraphics[width=0.5\textwidth]{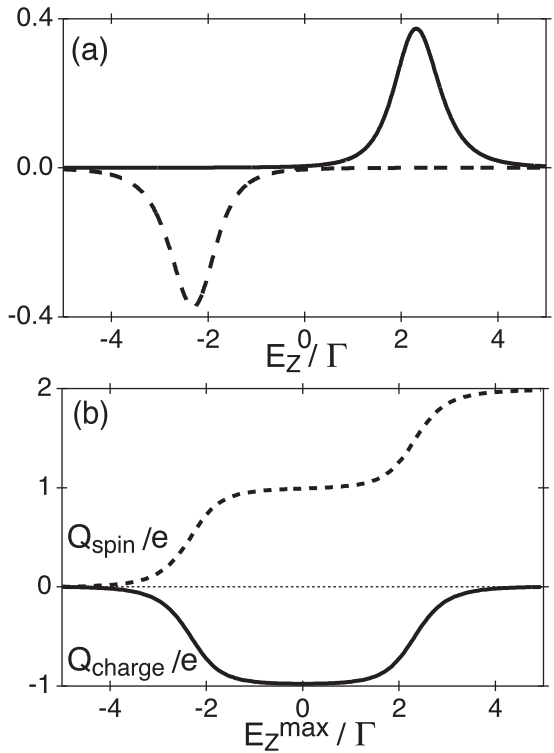}
\caption{\label{fig:pi} 
(a) Plots of $\Pi_{\sigma}$ versus $E_Z/\Gamma$:
Thick solid (broken) line represents
$\Pi_{+}$( $\Pi_{-}$) when
$U/\Gamma=10$ and $E_0/\Gamma=2.0$.
(b)Pumped charge, $Q_{\rm charge}=Q_{+}+Q_{-}$ and pumped spin, $Q_{\rm spin}=Q_{+}-Q{-}$ as
a function of $E_Z^{\rm max}/\Gamma$ with a fixed value of $E_Z^{\rm min}/\Gamma =-5$.
}
\end{figure}

Besides the spin pumping without the charge pumping,
we can control $Q_{\rm charge}$ and $Q_{\rm spin}$ flexibly
by changing the range of $E_Z$.
We illustrate this point by
calculating $Q_{\rm charge}$ and $Q_{\rm spin}$ using the mean field approximation (\ref{eq:MFA})
for the non-localized moment regime.
In Fig.~\ref{fig:pi}(a),
$\Pi_{+}(E_Z)$ and $\Pi_{-}(E_Z)$ are plotted 
when $U/\Gamma=10$ and $E_0/\Gamma =2$, and
in Fig.~\ref{fig:pi}(b),
 $Q_{\rm charge}$ and $Q_{\rm spin}$ are plotted as a function of
$E_{Z}^{\rm max}/\Gamma$ with a fixed value of
$E_{Z}^{\rm min}/\Gamma = -5$.
When $E_{Z}^{\rm max}$ is small enough,
both of $Q_{\rm charge}$ and $Q_{\rm spin}$ are zero.
As $E_Z^{\rm max}$ increases,
$Q_{\rm charge}$ and $Q_{\rm spin}$ become finite with opposite signs,
where only down spin charge flows from the right to left lead.
When $-2 < E_{Z}^{\rm max} /\Gamma < 2$,
both of $Q_{\rm spin}$ and $Q_{\rm charge}$ take plateau values
whose magnitudes are equal to $e$ though the signs are opposite.
When $E_{Z}^{\rm max}$ increases further,
$Q_{\rm charge}$ decreases to zero while
$Q_{\rm spin}$ increases to $2 e$, where
up-spin charge starts to flow in the opposite direction,
from the left to right lead, eventually canceling
down-spin charge to achieve
the spin pumping without charge pumping.
In the same way,
we can obtain other possibilities of $Q_{\rm spin}$ and $Q_{\rm charge}$,
choosing $E_Z^{\rm min}$ and $E_Z^{\rm max}$ appropriately.

Finally we compare the spin pumping model
with an other spin current generator
using a quantum dot
under a finite source-drain voltage and
a stationary magnetic field.
Since the Zeeman splitting of the dot level acts as a spin filter,
we can control the spin flow through the quantum dot
by choosing the source-drain voltage appropriately.
In this spin filter, however,
the up and down spin currents always flow in the same direction.
Thus spin current without charge current cannot be achieved.
The magnitude of the current of this spin filter is
on the order of
$e \Gamma$,
which is  much larger than
the one of
the spin pumping,
which is on the order of $e \omega$.

In conclusion,
we investigate an adiabatic spin pumping
through a  quantum dot with a single orbital level
using the Zeeman effect.
The maximum value of the unit charge is transferred per cycle.
The maximum value is suppressed when
the localized moment in the dot appears.
Pumped charge and spin are different, and
under a certain condition,
a spin is pumped with vanishing charge pumping.
They are
tunable simply by changing
the amplitude of magnetic fields.
This may introduce flexibility of spin and charge control
in semiconductor nano structures.

The author thanks
Prof.~Y.~Ono for fruitful discussions and comments.
He thanks Prof.~Y.~Avishai for useful comments.
He also thanks Profs.~O.~Entin-Wohlman and A.~Aharony for
fruitful discussions.
This work is supported by
JSPS Research Fellowships for Young Scientists.


\begin{thebibliography}{99}

\bibitem{Thouless83}
    D.J. Thouless, Phys. Rev. B {\bf 27}, 6083 (1983).

\bibitem{Kouwenhoven91} 
      L. P. Kouwenhoven, A. T. Johnson, N. C. van der Vaart,
      C. J. P. M. Harmans, and C. T. Foxon, 
       Phys. Rev. Lett. {\bf 67}, 1626 (1991).
\bibitem{Pothier92} 
      H. Pothier, P. Lafarge, C. Urbina, D. Esteve, and M. H. Devoret, 
      Europhys. Lett.\ {\bf 17}, 249 (1992).
\bibitem{Keller98}
M.~W.~Keller, J.~M.~Martinis, and R.~L.~Kautz,
Phys.~Rev.~Lett. {\bf 80}, 4530 (1998).

\bibitem{Switkes99} 
M. Switkes, C. M. Marcus, K. Campman, and A. C. Gossard,
Science {\bf 283}, 1905 (1999).
\bibitem{Spivak95} 
B.~Spivak, F.~Zhou, and M.~T.~Beal Monod,
Phys.~Rev.~B {\bf 51}, 13226  (1995).

\bibitem{Aleiner98}
I.~L.~Aleiner, A.~V.~Andreev,
Phys.~Rev.~Lett. {\bf 81}, 1286 (1998).

\bibitem{Zhou99} 
F. Zhou, B. Spivak, and B. L. Altshuler,
Phys. Rev. Lett. {\bf 82}, 608 (1999).

\bibitem{Brouwer98}
P.~W.~Brouwer,
Phys. Rev. B {\bf 58}, R10135 (1998).

\bibitem{Brouwer98_2}
P. W. Brouwer,
Phys. Rev. B {\bf 58}, 10135 (1998).    

\bibitem{Shutenko00} 
      T. A. Shutenko, I. L. Aleiner, and B. L. Altshuler,
      Phys.~Rev.~B {\bf 61}, 10366 (2000).
      
\bibitem{Avron00}
       J. E. Avron, A. Elgart, G. M. Graf, and L. Sadun, 
       Phys.~Rev.~B {\bf 62}, 10618 (2000).
       
\bibitem{Aleiner00}
I. L. Aleiner, B. L. Altshuler, and A. Kamenev,
  Phys.~Rev.~B {\bf 62}, 10373 (2000).     
  
\bibitem{Polianski01} M. L. Polianski and P. W. Brouwer, Phys. Rev. B
{\bf 64}, 75304 (2001).

\bibitem{Sharma01}
P.~Sharma and C.~Chamon,
Phys.~Rev.~Lett. {\bf 87}, 096401 (2001).

\bibitem{Moskalets01} M. Moskalets and M. B\"uttiker,
Phys.~Rev.~B {\bf 64}, 201305(R) (2001).

\bibitem{Cremers01}
J.~N.~H.~J.~Cremers, and P.~W.~Brouwer,
Phys.~Rev.~B {\bf 65}, 115333 (2002).


\bibitem{Simon00}
S. H. Simon, Phys. Rev. B {\bf 61}, R16327 (2000).

\bibitem{Wei00}
    Yadong Wei, Jian Wang, and Hong Guo,
    Phys.~Rev.~B {\bf 62}, 9947 (2000).

\bibitem{Levinson00}
    Y.~Levinson, O.~Entin-Wohlman, and P.~W\"olfle, 
Physica A {\bf 302}, 335 (2000).


\bibitem{Mucciolo01}
E.~R.~Mucciolo, C.~Chamon, and C.~M.~Marcus,
Phys.~Rev.~Lett. {\bf 89}, 146802 (2002).

\bibitem{Entin02}
    O.~Entin-Wohlman, and A.~Aharony, Phys.~Rev.~B {\bf 66}, 035329 (2002).
  

\bibitem{Lee99}
W.~Y.~Lee, S.~Gardelis, B.~C.~Choi, Y.~B.~Xu, C.~G.~Smith,
C.~H.~W.~Barnes, D.~A.~Ritchie, E.~H.~Linfield, and J.~A.~C.~Bland,
J.~Appl.~Phys. {\bf 85}, 6682 (1999).

\bibitem{Hammar99}
P.~R.~Hammar, B.~R.~Bennett, M.~J.~Yang, and M.~Johnson,
Phys.~Rev.~Lett. {\bf 83}, 203 (1999).

\bibitem{Fiederling99}
R.~Fiederling, M.~Keim, G.~Reuscher, W.~Ossau, G.~Schmidt,
A.~Waag, and L.~W.~Molenkamp,
Nature (London) {\bf 402}, 787 (1999).

\bibitem{Ohno99}
Y.~Ohno, D.~K.~Young, B.~Beschoten, F.~Matsukura, H.~Ohno, and D.~D.~Awschalom,
Nature (London) {\bf 402}, 790 (1999).


\bibitem{Zeemaninlead}  
If the Zeeman effect in the leads is taken into account,
the level broadening $\Gamma$ becomes spin dependent:
$\Gamma_{\sigma} = \pi \rho_{\sigma} (|V_L|^2 + |V_R|^2)$,
where $\rho_{\sigma}$ is the density of states at $\epsilon = \sigma \mu g_B B$.
If we assume the lead is a 2D electron gas,
which is a reasonable assumption,
$\rho$ is a constant.
Then
the spin dependence of $\Gamma$ does not appear.

\bibitem{Glazman88}
 L.~I. Glazman and M.~{\'E}. Ra{\u\i}kh,
Pis'ma Zh. Eksp. Teor. Fiz. {\bf 47}, 378 (1988).
[{JETP} Lett. {\bf 47},452 (1988).]

\bibitem{Ng88}
    T.~K.~Ng and P.~A.~Lee, Phys.~Rev.~Lett.~{\bf 61}, 1768 (1988).


\bibitem{Anderson61}
P.~W.~Anderson, Phys.~Rev.~{\bf 124}, 41 (1961).

      
\bibitem{Kaminski99} A.~Kaminski, Yu.~V.~Nazarov, and L.~I.~Glazman, Phys.~Rev.~Lett.
{\bf 83}, 384 (1999);
Phy.~Rev. B {\bf 62}, 8154 (2000).

 
\bibitem{Buttiker94}
    M. B\"uttiker, H. Thomas, and A. Pr\'etre, Z. Phys. B {\bf 94}, 133 
(1994).

\bibitem{Langer61}
J.~S.~Langer and V.~Ambegaokar,
Phy.~Rev.~ {\bf 121}, 1090 (1961).

\bibitem{Langreth66}
D.~Langreth,
Phys.~Rev. {\bf 150}, 516 (1966).

\end{thebibliography}
\end{document}